\documentclass{aa}
\usepackage{graphicx}
\begin{document}
   \title{The   nuclear  radio--optical  properties   of  intermediate
   redshift   FR~II  radio   galaxies   and  quasars\thanks{Based   on
   observation  obtained  at the  Space  Telescope Science  Institute,
   which is  operated by the Association of  Universities for Research
   in Astronomy, Incorporated, under NASA contract NAS 5-26555.} }

\author{S.   Varano  
        \inst{1}\fnmsep\thanks{Visiting  student,  Summer
Student    Program    2002}    
\and    M.    Chiaberge\inst{1}    
\and
F.~D.   Macchetto\inst{2}\fnmsep\thanks{Space Telescope Division, ESA}   
\and
A. Capetti\inst{3} 
}

\offprints{S. Varano \email{varano\_s@ira.cnr.it}}

\institute{Istituto  di  Radioastronomia  del  CNR, Via P. Gobetti 101, Bologna, I-40129 \\   
\and Space Telescope Science Institute, 3700 San Martin Dr., Baltimore, MD 21210 \\ 
\and INAF--Osservatorio Astronomico di Torino, Strada Osservatorio 20, Pino Torinese, I-10025\\ 
}
 
\date{Received....; accepted....}

\abstract {We  extend the study of  the nuclei of 3CR  objects as seen
with the HST to higher redshift FR II radio sources ($0.4<z<0.6$). Our
results reflect what has been found  for FR II of lower redshift.  The
position  of the  nuclei  on the  plane  formed by  optical and  radio
luminosities is related to  their optical spectral classification: the
nuclei of both high and low excitation galaxies lie on the correlation
found for FR~I sources,  while broad--lined objects have a significant
optical excess.   The nuclear properties  of these sources  are better
understood  when  considering  the  equivalent  width  of  the  [OIII]
emission line with respect to their optical luminosities.  Even in the
range  of  redshift  considered  here, low  excitation  galaxies  show
peculiar nuclear  properties, more similar to those  observed in FR~I.
This  confirms that  not  all  FR~II are  unified  with quasars.   Our
findings have important implications for the FR~II--quasar unification
scheme: by reconsidering the  classification of all 3CR radio galaxies
with $z<1$  in the  light of their  nuclear properties, we  derive the
opening angle of the obscuring  torus for different redshift bins.  We
find that the covering factor  of the obscuring structure decreases as
redshift   increases  ($\theta   \sim  44^{\circ}$   for   $z<0.3$  to
$\theta\sim 56^{\circ}$  for $0.5<z<1$).  We argue  that this behavior
may be  interpreted in the framework  of the receding  torus model, in
which the  opening angle of the  torus increases as  the luminosity of
the  accretion   disk  around   the  central  black   hole  increases.

}

\authorrunning{S. Varano et al.}
\titlerunning{Intermediate $z$ FR~II}
\maketitle
%
 
\section{Introduction} 
 
In the  framework of the radio--loud AGN  unification scheme, Fanaroff
\&  Riley I  and  II  (FR I  and  FR II)  radio  sources (Fanaroff  \&
Riley~\cite{fanaroffriley}) are  believed to be  the parent population
of BL Lacs and QSO, respectively (see Urry \& Padovani~\cite{urry} for
a  review).  In this  scenario, the  differences between  the observed
emission  properties are ascribed  to a  different orientation  of the
anisotropic  central  emission.   There   are  two  main  reasons  for
anisotropy: the presence of  non--thermal emission from a relativistic
jet,  which dominates  the  radiation  observed in  BL  Lacs and  Flat
Spectrum Radio  Quasars, and the  presence of an  obscuring ``torus'',
which prevents us  to observe directly the central  parsecs in objects
observed along lines--of--sight perpendicular to the jet axis.

Optical studies of radio  galaxies are particularly important not only
to investigate their large--scale properties, such as the structure of
the      host       and      the      environment       (see      e.g.
Zirbel~\cite{zirbel96},~\cite{zirbel97}),  but  also  to  explore  the
physical conditions  of the central regions.  The  high resolution and
sensitivity provided  by HST optical  images allow us to  separate the
AGN emission from the stellar  host galaxy background even in the case
of low power radio galaxies (FR~I).

Chiaberge et  al.~(\cite{pap1}) (hereafter, Paper~I)  and Chiaberge et
al.~(\cite{pap2}) (hereafter, Paper~II)  have studied complete samples
of FR~I and  FR~II radio galaxies with $z<0.3$  using HST images taken
as   part   of   the   3CR   snapshot   surveys   (e.g.    Martel   et
al.~\cite{martel99}; De  Koff et al.~\cite{dekoff96}).   They analyzed
the  properties of  unresolved  nuclei  which have  been  found to  be
present in the  great majority of the objects.   The optical nuclei of
FR~I  correlate linearly  with the  radio nuclear  emission  over four
orders of magnitude, both in flux and luminosity.  In analogy with the
origin of the radio cores, the CCC (central compact core) emission has
been interpreted  as non--thermal synchrotron radiation  from the base
of the relativistic jet.   Furthermore, the high detection rate (85\%)
suggests that  we have a  direct view of  the central regions  in most
objects, and that a  geometrically and optically thick obscuring torus
is generally  not present in  FR~I objects.  Thus,  FR~I intrinsically
lack strong broad emission lines.

FR~II  with redshift $z<0.3$  show a  more complex  behavior, although
their  properties  are  clearly  related  to  their  optical  spectral
classification.  Objects with  broad emission lines (broad--line radio
galaxies and quasars) have bright  nuclei which show an optical excess
with  respect  to  the  FR~I  correlation,  probably  because  of  the
dominance of thermal  emission from the accretion disk  in the optical
band. Sources in which only narrow lines are present are distinguished
into  high  and  low excitation  galaxies  (HEG  and  LEG -  Laing  et
al.~\cite{laing94}).  Approximately  50\% of the  HEG do not  show any
nuclear  optical  component  and  they  are  interpreted  as  obscured
sources, as expected  in the framework of the  unification scheme.  On
the  other hand,  the  nuclei of  HEG,  when present,  do  lie on  the
correlation (or slightly above  it).  This behavior is unexpected, and
their  nature  is  intriguing.   However,  by  including  the  nuclear
Equivalent Width of the [OIII] emission line as a further parameter in
the diagnostic  diagrams, we can discriminate  between nuclei observed
directly and those  that are obscured and seen  only through scattered
emission (Paper~II).

LEG seem  to behave differently  from other radio galaxies  with FR~II
morphology.  A significant fraction  of them have faint optical nuclei
similar  to  those  of  FR~I   and  they  lie  on  the  radio--optical
correlation.   Their  nuclear physical  properties  are probably  more
similar to those of FR~I, and  this appears to be in contrast with the
``standard'' zeroth--order unification  model.  These sources might be
associated with those  BL Lacs that show radio  morphologies and total
radio power more typical of  FR~II sources, thus providing support for
a revised  unification scenario (Laing  et al.~\cite{laing94}, Jackson
\& Wall~\cite{jacksonwall}, Hardcastle et al.~\cite{hard04}).

The limitation so  far has been that the  results described above hold
for low  redshift sources.   The aim  of this paper  is to  extend the
previous studies  to a sample of  higher redshift FR II,  in the range
$0.4<z<0.6$.  This  is particularly important for two  reasons: i) the
radio--loud AGN unification scheme  was originally proposed by Barthel
et al.~(\cite{barthel89}) using a  sample of high redshift ($0.5<z<1$)
3CR objects;  ii) cosmological evolution  may play a role  in changing
the  structure   of  radio   galaxies  and  quasar.    Therefore,  the
conclusions drawn  for the low redshift  sample may not  apply to high
redshift objects,  and the  discrepancies with the  unification scheme
found in the  analysis of nearby samples may be  explained in terms of
cosmological  evolution.   Instead,  if  the same  behavior  is  still
observed  as redshift increases,  the unification  scheme needs  to be
revised.

The paper is organized as follows: in Section 2 we describe our sample
of  FR~II and  in Section  3  the HST  observations; in  Section 4  we
present the results  of the photometry of the nuclei;  in Section 5 we
discuss  our   results  and   consider  their  implications   for  the
unification models.  In  Section 6, we summarize the  main results and
draw  conclusions  and future  perspectives.  ${\rm H_0=75\,km  s^{-1}
Mpc^{-1}}$ and $\rm{q}_0=0.5$ are used hereafter.
 
\section{The sample} 
 
\begin{table*}
\begin{tabular}{llrccc} 
\hline  Source   & Redshift  &   $\log{\rm L_{\rm r}}$(5GHz)  &   
$\log{\rm L_{\rm tot}}$(178MHz) & $\log{\rm L_{\rm [OIII]}}$ & Spectral\\ 
name & $z$  & ${\rm erg\,s^{-1}\,Hz^{-1}}$ &   
${\rm erg\,s^{-1}\,Hz^{-1}}$ & W & class.\\  
\hline   
\object{3C~16} & $0.405$  & $<30.45$ & $34.77$ & $-$ & HEG\\  
\object{3C~19} & $0.482$ & $<33.22$ & $35.00$ & $-$ & LEG\\  
\object{3C~46} & $0.4373$ &  $31.17$ & $34.82$ & $36.04$ & HEG\\  
\object{3C~47}  & $0.425$  & $32.65$  & $35.20$  & $36.52$ & QSO\\   
\object{3C~99} & $0.426$ &  $32.99$ & $34.81$  & $-$ & HEG$^b$\\   
\object{3C~147} &  $0.545$ & $34.22$ & $35.25$ & $37.03$ & QSO\\  
\object{3C~154} & $0.5804$ & $33.86$  &  $35.49$ & $-$ & QSO$^b$\\  
\object{3C~172} &  $0.5191$ &  $<31.70$ & $35.18$  & $-$ & HEG\\   
\object{3C~200}  & $0.458$  & $32.41$  & $35.01$  & $-$ & LEG\\  
\object{3C~215} &  $0.411$ &  $31.96$ &  $34.80$ & $35.83$ & QSO\\  
\object{3C~225.0B} & $0.58$ &  $<31.12$ & $35.45$ & $-$ & HEG?\\  
\object{3C~228}  & $0.5524$ &  $32.19$ &  $35.32$ &  $-$ & HEG\\  
\object{3C~244.1} & $0.428$ & $<30.56$ & $35.13$ & $36.27$ & HEG\\  
\object{3C~274.1} & $0.422$ & $31.60$ & $34.99$  & $34.60$ & HEG(LEG$^*$)\\  
\object{3C~275} & $0.48$ & $-$ & $35.09$ & $36.26$ & LEG$^a$\\ 
\object{3C~275.1}& $0.557$ &  $33.37$ & $35.34$ &  $35.81$ & QSO\\  
\object{3C~295} &  $0.4614$ &  $31.71$ & $34.79$ & $35.23$ & LEG\\ 
\object{3C~306.1} & $0.441$ & $-$ & $35.95$ & $-$ & HEG$^a$\\  
\object{3C~313} &  $0.461$ & $<30.87$ & $35.17$ & $35.71$ & HEG$^a$\\  
\object{3C~327.1} & $0.4628$  & $32.68$ & $35.24$ &  $35.95$ & HEG$^a$\\
\object{3C~330} & $0.55$ & $30.93$ & $35.51$ & $-$ & HEG\\  
\object{3C~334} & $0.5551$ & $33.12$ & $35.11$ &  $36.61$ & QSO\\  
\object{3C~341} & $0.448$  & $30.84$ & $34.87$  & $36.04$ & HEG\\  
\object{3C~345} &  $0.594$ & $34.87$ & $35.23$ &  $36.17$ & QSO$^a$\\  
\object{3C~411} &  $0.467$ &  $32.53$ & $35.10$ &  $-$ & HEG$^a$\\  
\object{3C~427.1} & $0.572$  & $30.57$ & $35.53$ & $-$ & LEG\\ 
\object{3C~435A} & $0.471$ & $32.12$ & $34.95$ & $-$ & Unclass\\
\object{3C~455} & $0.543$ & $31.19$ & $35.19$ & $36.31$ & WQ\\    
\hline
\end{tabular} 
\caption[]{Summary of  data from the  literature for our  sample.  The
spectral  classification   is  from  Willott's  catalog   at  the  URL
http://www-astro.physics.ox.ac.uk/~cjw/3crr/3crr.html,  except (a) for
which   we   used  the   classification   published   in  Jackson   \&
Rawlings~(\cite{jacksonrawlings})     (spectra    by     Jackson    \&
Browne~\cite{jacksonbrowne}  for  \object{3C~345},  Tadhunter  et
al.~\cite{tadhunter93}  for  \object{3C~327.1}  and Spinrad  et
al.~(\cite{spinrad}) for  the rest)  and (b) for  which we  report the
classification available  in the NASA/IPAC  Extragalactic Database (no
references   for  the   spectra).\\$^*$   \object{3C~274.1}  is
classified by Willott as a HEG, but Jackson \& Rawlings classify it as
a doubtful LEG.  See discussion in Sect. \ref{sez:oiii}.}
\end{table*} 

The  sample considered  in  this paper  comprises  all radio  galaxies
belonging  to the  3CR  catalog (Spinrad  et al.~\cite{spinrad})  with
redshift in  the range  $0.4<z<0.6$, morphologically classified  as FR
II.
 
We checked  the morphological classification,  optical identification,
and redshift of all sources,  by searching the literature for the most
recent data.   We exclude  \object{3C~119} from our  sample since
its  redshift  is  $z=1.023$  (Eracleous  \&  Halpern~\cite{eracle94})
instead of $z=0.408$, as originally reported in the 3CR catalog.
The final list is thus a complete, flux and redshift limited sample of
28  FR II radio  sources (Table  1).  Note  that, to  the best  of our
knowledge, for \object{3C~306.1} and \object{3C~275} no information on
the radio core is available .
 
We  classify  our  sources  on   the  basis  of  their  emission  line
properties,   adopting    the   scheme   defined    by   Jackson   and
Rawlings~(\cite{jacksonrawlings}).   Weak  Quasars  (WQ)  and  quasars
(QSO) are objects for which at least one broad line has been observed.
These two  classes differ only in optical  continuum luminosity: ${\rm
L_{\rm 5500 \AA}}>10^{23}$ W Hz$^{-1}$  for QSO, while WQ are fainter.
Thus, in  the following, we  refer to them  as a single  population of
broad lined objects (BLO).  High  and low excitation galaxies (HEG and
LEG) are narrow--lined sources, and  are distinguished on the basis of
their [OII]/[OIII]  ratio ($>{1}$ for a LEG)  and/or [OIII] equivalent
width (less than $10$\AA~for a LEG).
 
In Table 1 we report radio and optical data for the complete sample as
taken from the literature .

\begin{table}[b] 
\begin{tabular}{lccc} 
\hline Source Name & Filter & ${\rm t_{\rm exp}}$(s) & GO program ID \\ 
\hline 

\object{3C~16} & F702W & $8300$ & $6675$\\
\object{3C~19} & F785LP & $1000$ & $9045$\\
\object{3C~46} & F702W & $300$ & $5476$\\
\object{3C~47}  & F702W & $280$ & $5476$\\
\object{3c~99} & F702W & $300$ & $5476$\\ 
\object{3C~147} & F702W  & $280$ & $5476$\\  
\object{3C~154} &  F702W & $280$ &  $5476$\\ 
\object{3C~172} & F702W  & $300$ & $5476$\\  
\object{3C~200} & F702W &  $300$ & $5476$\\ 
\object{3C~215}  & F702W &  $280$ &  $5476$\\ 
\object{3C~225.0B}  & F702W &  $300$ & $5476$\\ 
\object{3C~228} &  F702W & $300$  & $5476$\\  
\object{3C~244.1} &  F702W & $300$ & $5476$\\ 
\object{3C~274.1} &  F785LP & $2000$ & $9045$\\
\object{3C~275} & F702W & $300$& $5476$\\ 
\object{3C~275.1} & F675W &  $1000$ & $5978$\\  
\object{3C~295} &  F702W & $6300$ &  $5378$\\
\object{3C~306.1} & F702W & $300$ & $5476$\\ 
\object{3C~313} & F702W  & $300$ & $5476$\\  
\object{3C~330} & F555W &  $600$ & $6348$\\ 
\object{3C~334} &  F702W &  $280$ &  $5476$\\  
\object{3C~341}  & F702W  & $300$  & $5476$\\ 
\object{3C~345} & F555W & $2800$ & $5235$\\ 
\object{3C~411} & F702W & $300$ & $5476$\\ 
\object{3C~427.1} & F702W &  $300$ & $5476$\\ 
\object{3C~435A} & F702W & $7400$ & $6675$\\ 
\object{3C~455}  & F702W & $300$ & $5476$\\ 
\hline 
\end{tabular} 
\caption[]{Log of HST observations.} 
\end{table}

\section{HST observations}\label{sez:hstobs} 
 
Optical HST observations  are available in the public  archive for all
but one (namely \object{3C~327.1}) of the sources of our sample .  The
HST  images were taken  using the  Wide Field  and Planetary  Camera 2
(WFPC2).  The  projected pixel size  of the Planetary Camera  (PC) and
the  Wide Field  cameras (WF)  are $0.0455$  arcsec and  $0.1$ arcsec,
respectively.  The three WF cameras cover a ``L'' shaped field of view
of $150^{\prime\prime} \times 150^{\prime\prime}$, while the PC covers
an area of $35^{\prime\prime} \times 35^{\prime\prime}$.
 
Twenty--two sources were observed with the F702W filter as part of the
HST  snapshot  survey  of  3CR  radio  sources  (GO~5476,  Lehnert  et
al.~\cite{lehnert99};     De     Koff     et     al.~\cite{dekoff96}).
\object{3C~295}  was observed  with  the  F702W filter  as  part of  a
different    program    (GO5378),    while    \object{3C~275.1}    and
\object{3C~345} were observed with  the F675W and F555W, respectively.
Deep  observations  with  the  F702W  filter are  also  available  for
\object{3C~16} and \object{3C~435} from GO~6675.  Further observations
with  the  F785LP  filter  (taken  as part  of  program  GO~9045)  are
available for  eleven radio galaxies of our  sample. However, although
the  F785LP filter has  a wide  passband ($\sim  2000$ \AA),  and deep
observations  were obtained  (${\rm  t_{\rm exp}}  =  2000$ sec),  the
target is always located in the WF3 Camera.  Therefore, since our goal
is  to check  for the  presence of  unresolved nuclear  components, we
favor observations  in which the  target lies in  the PC, which  has a
smaller projected pixel size, thus  providing a better sampling of the
PSF. Since \object{3C~327.1} has not been observed with HST, our final
sample is composed by twenty--seven sources.  In Table 2, we summarize
the details of the HST observations.

The  data  have  been  processed  through  the  standard  OTFR  system
(On--The--Fly   Reprocessing)  pipeline   for   data  processing   and
calibration (Baggett et al.~\cite{bagg02}).

For  twelve  sources  the  observations  are  splitted  into  multiple
exposures in order to remove cosmic  ray events.  We use the IRAF task
CRREJ which  combines the  images and rejects  cosmic rays  through an
iterative  process.   For fifteen  out  of  the twenty--three  sources
observed   with  the   F702W   filter  multiple   exposures  are   not
available. In these  cases, we carefully check the  central regions of
the target to ensure us that  the area of interest is not contaminated
by cosmic rays. Cosmic rays are  present in the nuclear regions of the
F702W     images    of    \object{3C~19},     \object{3C~274.1}    and
\object{3C~330}. Therefore, for such  objects we use observations with
different  filters (see Table  2). For  \object{3C~275}, a  cosmic ray
affects the central region of the only HST image available, preventing
us from any estimate of the nuclear emission.

\begin{figure}[b!]
\centering\includegraphics[width=8cm]{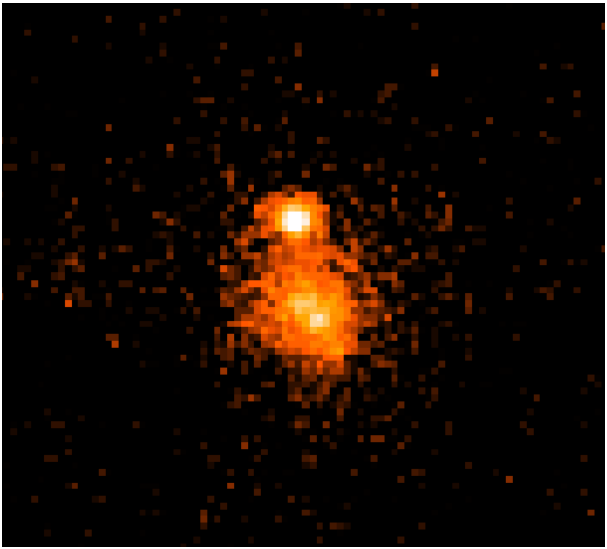} 
\caption{HST  image of  the  host galaxy  of  \object{3C~313} (in  the
center).  The  size of the  image is $4\times4\,{\rm  arcsec^2}$.  The
nucleus of this source has a complex morphology (see discussion in the
text, Sect.  \ref{sez:ident}).}
\end{figure} 

\section{Identification and photometry of Central Compact Cores}\label{sez:ident} 
 
The  identification   of  an  unresolved  nuclear   component  is  not
straightforward  for sources  in the  redshift range  considered here.
Typically, the host galaxies' diameter does not exceed $\sim10$ pixels
and a steep radial brightness profile might be confused with a nuclear
compact component. Thus, a  careful inspection of the galaxy's profile
is needed.

We use the IRAF  {\sl RADPROF} task on a region of  10 pixels from the
galaxy center  to derive its radial brightness  profile.  A steepening
of  the brightness  profile  occurring in  the  central pixels,  which
results in  a FWHM comparable to  that of an  unresolved source ($\sim
0.08$ arcsec) is indicative of the presence of a central compact core.

In twelve out of the  twenty--seven sources ($44\%$ of the sample) the
FWHM of the central region is $<0.08$ arcsec.  In order to measure the
flux of the CCC we perform aperture photometry.  The highest source of
error  is the  determination  of  the background:  we  set the  galaxy
background level at a distance of  four or five pixels from the center
of  the galaxy,  for  the objects  with  $0.5<z<0.6$ and  $0.4<z<0.5$,
respectively.

In seven galaxies we measure a  FWHM of the central region larger than
$0.1$ arcsec.   For these  objects we estimate  an upper limit  to the
nuclear source by measuring the light excess in the central $3\times3$
pixels with respect to the surrounding galaxy background.  For sources
in the redshift  range $0.5<z<0.6$ we lower the  central box size down
to $2\times2$ pixels.

Four  objects, namely  \object{3C~16},  \object{3C~46}, \object{3C~99}
and \object{3C~313},  show a complex  morphology which prevents  us to
detect  any nuclear  emission. Note  that the  optical  counterpart of
\object{3C~313} was previously identified on the HST image as a bright
(quasar--like)  point  source (De  Koff  et al.~\cite{dekoff96}).  Its
radio core  is only marginally detected  in a VLA  observation at 5GHz
(Baum et al.~\cite{baum88}),  but it is clearly seen  at 8 GHz (Bogers
et al.~\cite{bogers94}).  The position  of the core, as estimated from
the radio map,  corresponds to a faint galaxy on  the HST image, which
is located close to the edge  of the PC field (Fig.  1).  This object,
which  we consider  as the  true  host galaxy  of \object{3C~313},  is
located $\sim 5$  arcsec away from the previous  identification in the
HST  image by  De  Koff et  al.~(\cite{dekoff96}).  Unfortunately,  as
reported  above,  the  host  galaxy  of \object{3C~313}  has  a  close
companion  and  shows a  complex  nucleus,  with  possible dust  lanes
obscuring the central region.

A further  problem in estimating the nuclear  continuum is represented
by  contamination  from  emission  lines,  which  may  be  significant
especially in the case of BLO and HEG.  In Paper~II we have considered
the contamination from broad H$\alpha$  lines in BLO.  We showed that,
for  the range  of redshifts  in which  the line  falls in  the filter
passband, the  flux of the  emission line represents only  10--20\% of
the  nuclear flux.  For  the range  of redshifts  and for  the filters
considered in  this paper, the  emission lines that may  contribute to
the nuclear flux  observed in the broad band  filter images are mainly
represented   by   the  [OIII]5007   and   H$\beta$.    For  the   HEG
\object{3C~244.1}  an  HST  image  taken  with a  narrow  band  filter
(FR680N), centered on the [OIII]5007  line, is available in the public
archive.  It is well known that a significant fraction of the emission
line  flux in  radio  galaxies is  produced  in extended  (kpc--scale)
regions  of  the  host   (e.g.   Baum  et  al.   \cite{baum88}).   For
\object{3C~244.1},   we  estimate   that  the   [OIII]   nuclear  flux
corresponds  to $\sim  50\%$ of  the  total [OIII]  flux derived  from
ground      based      observations      (Jackson     \&      Rawlings
\cite{jacksonrawlings}). We have measured the nuclear [OIII] flux from
the narrow band image, and we  find that its contribution to the total
count--rate observed in  the broad band image, results  in $\sim 50\%$
of  the nuclear  flux.   Although  this appears  to  be a  substantial
contribution, it  does not significantly affect our  results.  For two
more  HEG  (namely  \object{3C274.1}  and \object{3C~341})  the  total
[OIII] flux is available from the literature, but their nuclei are not
detected in  our HST  images.  If  we assume that  $\sim 50\%$  of the
total [OIII] flux were produced in the nucleus, this would result in a
count-rate  higher than our  detection threshold  of $\sim  1.5 \times
10^{-29}$ erg cm$^{-2}$ s$^{-1}$ Hz$^{-1}$, as set by the upper limits
we estimate for these objects. This implies that most ($>50\%$) of the
[OIII]  flux  is produced  in  extended  (kpc--scale)  regions of  the
galaxy, and thus our estimates  for the nuclear continuum are suitable
for the purpose of the work presented here. However, note that even in
case the three CCC of the HEG were entirely line emission, our results
are not  substantially affected by substituting  detections with upper
limits, since  their representative points in  the diagnostic diagrams
would point to the regions were we expect the HEG to be located.

The results of our search for  CCC is summarized in  Table~3 and Table~4.

\begin{table}
\begin{tabular}{lr|lr} 
\hline  Source &  $\log{\rm L_o}$  (7000 $\AA$)  & Source  & $\log{\rm
L_o}$(7000 $\AA$)\\ name  & ${\rm  erg\,s^{-1} Hz^{-1}}$ &
name & ${\rm erg\,s^{-1} Hz^{-1}}$\\
  
\hline    
\object{3C~16} & Complex &  \object{3C~275}   & CR\\
\object{3C~19} & $<28.50$ & \object{3C~275.1} & $30.15$\\ 
\object{3C~46} & Complex &  \object{3C~295}   & $<27.67$\\ 	
\object{3C~47} & $30.20$ &  \object{3C~306.1} & $<28.20$\\	
\object{3C~99} & Complex &  \object{3C~313}   & Complex\\  	
\object{3C~147} & $31.01$ & \object{3C~327.1} & not obs.\\  
\object{3C~154} & $30.93$ & \object{3C~330} & $<27.85$\\  	
\object{3C~172} & $<27.29$ &\object{3C~334} & $31.04$\\  	
\object{3C~200} & $28.97$ & \object{3C~341} & $<28.29$\\  	
\object{3C~215} & $30.13$ & \object{3C~345} & $30.85$\\  	
\object{3C~225.0B} & $<27.69$ & \object{3C~411} & $29.52$\\ 
\object{3C~228} & $28.92$ & \object{3C~427.1} & $<26.99$\\ 
\object{3C~244.1} & $28.95$ & \object{3C~435A} & $<27.63$\\      
\object{3C~274.1} & $<28.26$ &\object{3C~455} & $29.88$\\    
\hline
\end{tabular} 
\caption[]{Result  from  identification   and  photometry  of  Central
Compact Cores in our sample.}
\end{table}

\section{Results and discussion} 

In order  to study the nature of  the nuclei we take  advantage of the
results obtained for the lower  redshift samples ($z<0.3$) in Papers I
and II.  In Figure  2 we plot our sources in the  plane defined by the
CCC luminosity vs the radio core luminosity and compare their behavior
to  that of  the nearby  3C  galaxies.  Note  that, due  to the  small
redshift  range of  our sample,  the distributions  in either  flux or
luminosity do not significantly differ.

The position of our sources on the luminosity plot with respect to the
FRI correlation  is related  to their optical  spectral classification
(Fig. 3).   Their behavior is  similar to that  of low redshift  FR II
sources, but shifted toward higher luminosities.

\begin{figure}[tb]   
\centering\includegraphics[width=8cm]{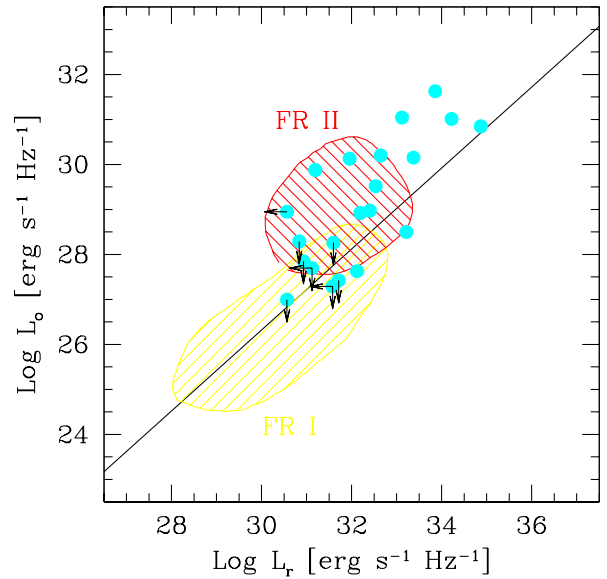}
\caption{Optical luminosity of the  CCC vs. radio core luminosity. The
two shaded regions represent the positions  of the FR I sample and the
sample of FR II at $z<0.3$.}
\end{figure}				 

Broad Lined Objects  are the brightest, both in  the optical and radio
band.  They typically show an  optical excess of 1--2 dex with respect
to  the FR  I correlation.   As discussed  in Paper  II,  this optical
excess  can be  explained  if  the optical  emission  is dominated  by
thermal  radiation  from  the  accretion disk.   However,  one  source
(\object{3C~345}) lies on the radio--optical correlation.  This is not
surprising, since \object{3C~345} is  a Flat Spectrum Radio Quasar and
both  its  radio  and   optical  nuclear  emission  are  dominated  by
non--thermal     synchrotron    radiation    (Moore     \&    Stockman
\cite{moorestock}). Note  that the apparent  correlation between ${\rm
L_{\rm o}}$ and ${\rm L_{\rm r}}$ for  BLO is not real and only due to
the common dependence on redshift  of the two quantities.  In fact, we
find that no correlation is present when considering fluxes instead of
luminosities.
  
The  nuclei  of the  LEG  lie  very close  to  the  FR I  correlation.
Although we  only have three sources  (one CCC and  two upper limits),
their behavior is similar to that of low--z LEG.

The HEG show a more complex behavior.  Six out of seven lie very close
to the FR I regression  line (within $2\sigma$). Instead, the location
of \object{3C~244.1} is puzzling.   Its representative point lies more
than  2 dex  out  of the  correlation.   Its optical  core is  clearly
identified in the HST image but  only an upper limit to the radio core
flux  is available.   We will  discuss in  detail the  nature  of this
object in the following section.

As shown in Paper~II, the position of HEG nuclei on the radio--optical
plane can be  misleading.  A significant fraction of  low redshift HEG
lie  on the FR~I  correlation ``by  chance'' and  the nature  of their
optical  emission can  be assessed  only when  a further  parameter is
included,  i.e. the  equivalent  width of  the  [OIII] emission  line.
Therefore, before  drawing conclusions on the nature  of high redshift
FR~II nuclei, we discuss the properties of our sources as far as their
emission lines are concerned.

\begin{figure}[tb]
\centering\includegraphics[width=8cm]{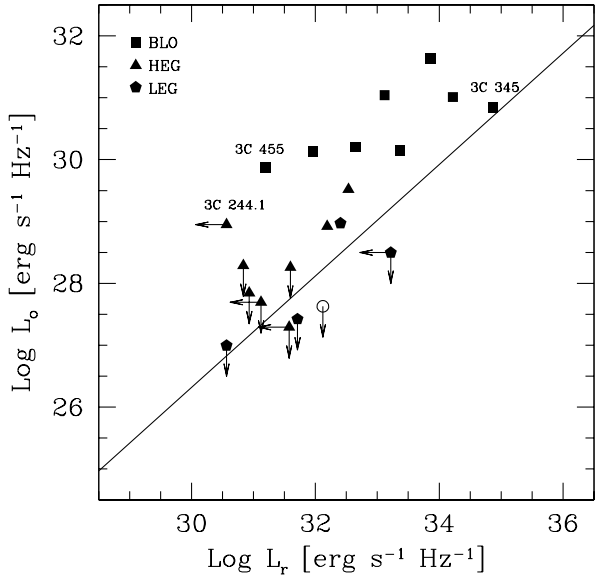} 
\caption{Same as Fig.  2; enlargement  of the range of luminosities of
the  present sample.   Different symbols  refer to  different spectral
classification. The empty circle  corresponds to the only unclassified
source (namely \object{3C~435A}).}
\end{figure} 

\subsection{Equivalent width of the [OIII] emission line vs. 
the radio--to--optical flux ratio}\label{sez:oiii}

In Figure  4 we plot the  ratio between the [OIII]  emission line flux
and the optical continuum (which represents the ``nuclear'' Equivalent
Width of [OIII]) vs the optical--to--radio core ratio. The position of
the sources along the Y axis indicates whether the nucleus is observed
directly or  it is  obscured to our  line--of--sight and  visible only
through scattered light.  In fact, in the scenario in which the narrow
line region is photo--ionized by  UV radiation produced in the central
accretion  disk, the  strength of  the emission  lines  (in particular
those  of  high ionization  levels)  is  closely  correlated with  the
intensity  of the  nuclear source.   If we  have a  clear view  of the
central regions of the galaxy, a low EW of the [OIII] emission line is
expected. Conversely, a  high EW ($\log$ EW$>3.5$) is  measured if the
nuclear (ionizing)  component is  obscured to our  line of  sight, and
only  a small  fraction of  it is  seen through  scattered  light (see
Paper~II for an extensive discussion).

The location  of the representative  points along the X  axis reflects
the distribution observed in the  plane formed by the optical vs radio
core emission.   The shaded region represents  the correlation between
these  two   quantities  found  for  FR~I   radio  sources  ($1\sigma$
dispersion).  Objects lying  on the right side of  this region have an
optical excess with respect to  the correlation. In the nearby sample,
BLO are  found in the lower right  region of the plane,  while the HEG
typically show  high EW and  are interpreted as obscured  BLO.  Low--z
LEG, all  of which lie  among the FR  I in all diagnostic  planes, are
interpreted as ``FRI--like'' nuclei and  they do not show evidence for
substantial obscuration.

Unfortunately, to  the best of  our knowledge, only eleven  sources of
our sample have optical  spectral information in the literature. Three
of them are HEG, one is a LEG and seven are BLO.

\begin{figure}[hb] 
\centering \includegraphics[width=7cm]{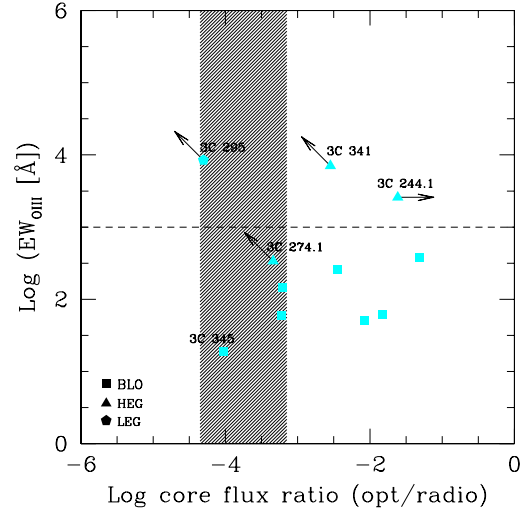} 
\caption{Equivalent Width  of the [OIII] emission  line, measured with
respect to the CCC emission, plotted vs. the ratio between the optical
CCC  to radio  core flux.  The shaded  area represents  the dispersion
($1\sigma$)  of the  linear  correlation between  radio  core and  CCC
luminosity found  for FR I in  Paper I. The dashed  horizontal line is
the  separation between ``obscured''  and ``unobscured''  sources (see
text, Sect. \ref{sez:oiii}).}
\end{figure} 

All of the BLO lie in the lower--right region of the plane in Figure 4
and have  $\rm EW_{\rm [OIII]}\sim100$.   This is a signature  that we
are  directly  observing the  source  of  ionization.   Note that,  as
already mentioned above, \object{3C~345}  is an outlier because of the
dominance of non--thermal emission in  the optical band.  In the plane
of Figure 4,  this has two effects: i)it drives  the object toward the
FR~I ``synchrotron'' correlation and ii)it lowers the equivalent width
of  [OIII],  as   the  emission  lines  are  diluted   by  the  strong
non--thermal continuum.

Two   out    of   the   three   HEG    (namely   \object{3C~341}   and
\object{3C~244.1}) have a high EW[OIII],  thus they are similar to the
vast majority of low redshift HEG, which lie in the top (left) part of
the diagram.   \object{3C~244.1} not only shows an  optical excess, as
already pointed  out in the previous  section, but it also  has a high
value of EW[OIII].  Although the non--detection for its  radio core is
puzzling,   its  behavior   appears   to  be   similar   to  that   of
\object{3C~184.1}  at $z=0.118$ (Paper  II).  Thus  we argue  that the
location  of  \object{3C~244.1}  in   the  diagnostic  planes  can  be
explained   if   this  object   has   an   unusually  high   intrinsic
optical--to--radio  core  flux ratio  and  its  ``quasar'' nucleus  is
moderately absorbed, so that its  optical broad emission lines are not
detected.  Alternatively, this  object may be a ``normal''  HEG with a
scattered  nuclear   continuum  (but  still  with   a  rather  unusual
optical--to--radio core  flux ratio).  To test  these hypothesis, both
deeper  radio data  and  better spectral  information  are needed:  to
confirm  the  former  hypothesis  we  should search  for  faint  broad
emission lines in  direct light, while for the  latter we expect broad
emission lines  (and nuclear continuum)  to be strongly  scattered and
thus polarized.

\object{3C~274.1} lies among the ``unobscured'' sources, where LEG are
usually found.   However, the non  detection of the optical  core only
sets a  limit to  its location on  this plane.  Its  [OIII] luminosity
($\log{\rm L_{\rm [OIII]}}=34.60$) is typical of LEG and significantly
smaller  than any  other HEG  in our  sample  ($\log{\rm L_{[OIII],\rm
med}}\sim{36.37}$).  Therefore, we argue  that this source is actually
a   LEG,   as   already    tentatively   proposed   by   Jackson   and
Rawlings~(\cite{jacksonrawlings}).

\begin{figure*}[ht] 
\centering \includegraphics[width=17cm]{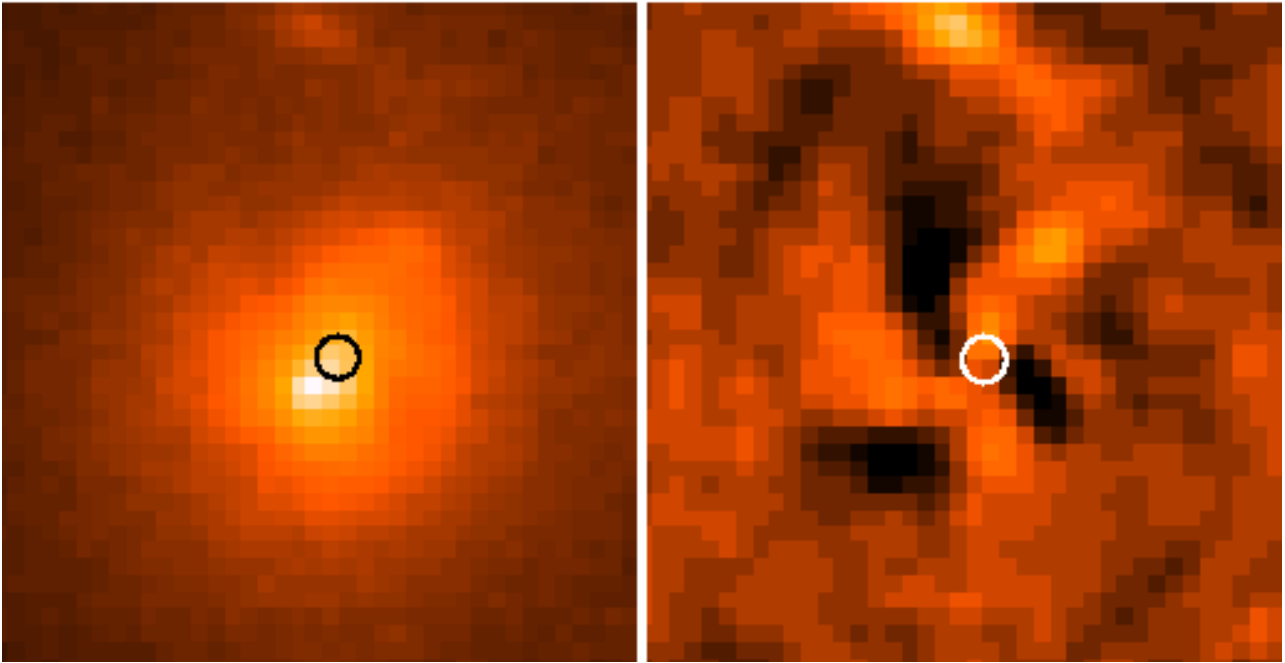} 
\caption{HST  optical  image  of  \object{3C~295} (on  the  left)  and
residuals after the subtraction of the model obtained with a isophotal
fitting  (on  the right).  The  circle is  the  center  of the  galaxy
model.  The dimension  of  both of  the  boxes is  $1.8\times1.8\,{\rm
arcsec^2}$.  See text for discussion (Sect. \ref{sez:oiii}).}
\end{figure*}

The only LEG with  [OIII] information, \object{3C~295}, lies among the
obscured sources, well out of the region occupied by low redshift LEG.
As  already pointed  out above,  low  redshift LEG  are usually  found
within the  FR I  region, i.e.  on  the radio-optical  correlation and
with low EW [OIII].

The     [OIII]     luminosity     of    \object{3C~295}     ($\log{\rm
L_{[OIII]}}=35.23$) is significantly higher than the average of all of
the 3C LEG, and  it is similar to that of HEG.   This could imply that
the source  has been  misclassified.  On the  other hand,  the optical
image (Fig.  5, left panel) reveals that the host galaxy has a complex
morphology and presents signs of interaction.  In order to test if the
lack of nuclear source is due  to the presence of dust that covers the
nuclear regions,  we perform  isophotal fitting of  the galaxy  and we
subtract  the   resulting  model  from  the   image.   This  procedure
highlights possible dishomogeneities in  the galaxy structure (Fig. 5,
right panel).  The resulting image confirms that the morphology of the
galaxy  is  complex.   There  is evidence  for  absorbing  structures,
possibly a dust lane, that  might lower the emission from the nucleus.
Furthermore, the center of the  galaxy lies at a distance of $\sim0.2$
arcsec  from   the  peak  of  its  surface   brightness  profile  (the
``nucleus'' for our estimate of the optical upper limit).  We conclude
that because  of the complexity of  the central region  of the galaxy,
the upper limit measured on  the HST image is probably unreliable, and
the EW([OIII]) is in turn overestimated.  An ${\rm A_{\rm V}}\sim 2-3$
mag is sufficient to hide the central source.  In light of the present
data, we cannot discriminate  between the two scenarios proposed, i.e.
a  misclassification of  the  source (which  would  be a  HEG) or  the
presence of  a moderate amount of absorption  produced by large--scale
dust lanes that prevents us to detect the optical nucleus.

Summarizing, the  behavior of the  high redshift BLO is  completely in
agreement with that  of their low redshift counterparts.   The lack of
complete spectral information  for both the LEG and  HEG in the high-z
sample  prevents  us from  drawing  any  firm  conclusion about  these
sources. However, we have shown that it is plausible that the scenario
proposed for  nearby objects  still fits the  behavior of  the distant
sample: LEG have FR~I-like nuclei and HEG are absorbed QSO.

\begin{table*} 
\begin{tabular}{lccccc} 
\hline  {}& CCC detections & Optical Upper  Limits  & Complex Morphology 
& TOT  & \% CCC detections\\  
\hline 
BLO& $8$ & $-$ & $-$ & $8$  & $100\%$\\ 
LEG& $1$ & $3$ & $1$  & $5$  & $20\%$\\  
HEG&  $3$ & $6$  & $4$  & $13$  & $21\%$\\
\hline
{\bf All sources} & $12$ & $9$ & $5$ & $26$ & $44\%$\\
\hline 
\end{tabular} 
\caption[]{Summary  of  the  optical  properties  of  the  sources  at
$0.4<z<0.6$  observed with HST and  results of our  study on the
images.  Note that  only one source of our  sample (classified as HEG)
has  no HST  data available  (\object{3C~327.1}).  \object{3C~275} has
also been excluded because of a cosmic ray event in the central region
of the HST image (see text, Sect.\ref{sez:hstobs}).}.
\end{table*}

\subsection{Implications for the Unified Models} 
 
Let us  now compare our findings  with the behavior  of lower redshift
FR~II radio  sources in  the framework of  the unification  models for
quasars    and    narrow-lined     radio    galaxies    (Barthel    et
al.~\cite{barthel89}).

In the  range $0.4<z<0.6$, 28\% of  the sources are BLO,  50\% are HEG
and 18\% are  LEG (only one object is  unclassified).  Compared to the
low redshift  sample of Paper~II (65  objects, of which  20\% are BLO,
46\% are HEG,  25\% are LEG and 9\% are  unclassified) the fraction of
BLO increases with  redshift while the fraction of  LEG decreases.  On
the other hand, the fraction of HEG appears to be fairly constant.  If
we translate  this result into  an estimate for the  ``opening angle''
$\theta$ of the obscuring torus, using ${\rm P=1-\cos\theta}$ (where P
is the fraction of objects  with broad lines), for a randomly oriented
sample of sources we obtain $\theta=44^{\circ}$ for $0.4<z<0.6$, while
for $z<0.3$ we have $\theta=37^{\circ}$.

However, according to the new picture presented here, in Paper~II, and
in   agreement    with   the    previous   findings   of    Laing   et
al.~(\cite{laing94}), it  is plausible that LEG  constitute a separate
sample,  and they  are not  to be  included in  the statistics  of the
unification scheme  for FR~II  and quasars.  In  this case,  we obtain
$\theta=49^{\circ}$   for   $0.4<z<0.6$   and   $\theta=43^\circ$   or
$46^{\circ}$ for $z<0.3$, considering  all the unclassified objects as
HEG or LEG, respectively.

Barthel  et  al.~(\cite{barthel89})  found $\theta=44^{\circ}$ as  the
dividing  angle between QSO  and radio galaxies for  the sample of 3CR
sources with $0.5<z<1$.  However, if we re-classify the sources in the
light of their  nuclear  properties and  considering the most   recent
spectral
information\footnote{http://www-astro.physics.ox.ac.uk/~cjw/3crr/3crr.html},
we obtain that in the Barthel sample 44\% of the objects are BLO, 53\%
are HEG and  only 2\%  (corresponding  to  1  source) are  LEG.   This
corresponds to a torus opening    angle of $56^{\circ}$.    Therefore,
regardless  of the shape  of the obscuring structure,  this is a clear
indication for a decreasing covering factor with increasing redshift.

Willott et al.~(\cite{willott00}) have  found that the quasar fraction
(and thus the  torus opening angle) in low  frequency selected samples
of radio sources increases  as luminosity and redshift increase.  They
proposed  that the statistical  properties of  their samples  could be
explained  by the  receding torus  model (Lawrence~\cite{lawrence91}).
In this  scenario, the  internal radius of  the obscuring  dusty torus
increases as the  luminosity of the central AGN  increases, since dust
sublimates  at  a fixed  temperature  (${\rm  T}\sim  1500$K).  As  an
alternative, they pointed out that the rising of a population of low-z
FR~I-like ``starved quasars'' can also account for their findings.

By classifying the objects  according to their nuclear properties, and
thus considering the LEG as  a separate population, we still find that
the opening angle increases  with redshift. Therefore, having excluded
the majority of the FR~I-like  population, we favor the receding torus
model as  the most plausible  scenario to account statistics  of FR~II
and quasars.

\begin{table} 
\begin{tabular}{lccc} 
\hline  {}& $z<0.3$ & $0.4<z<0.6$  & $0.5<z<1$ \\  
\hline 
BLO&      $20\%$ & $28\%$ & $44\%$ \\ 
HEG&      $46\%$ & $50\%$ & $53\%$ \\  
LEG&      $25\%$ & $18\%$ & $2\%$  \\
Unclass.& $9\% $ & $ 4\%$ & $1\%$  \\
\hline 
\end{tabular} 
\caption[]{Fraction of BLO, HEG and LEG for different bins of redshift. The low redshift bin corresponds to the sample of Paper~II, the intermediate to this paper, and the high-z bin corresponds to the sample of Barthel \cite{barthel89}.}
\end{table}

%
%
%
%
 
\section{Conclusions}

We have  analyzed the  optical emission  from the nuclei  of 26  FR II
radio sources within the  redshift range $0.4<z<0.6$.  Our results are
consistent  with  the behavior  observed  in  lower redshift  objects.
Although in this  range of redshifts the identification  of CCC is not
straightforward, $\sim$50\% of the  sources show an unresolved nuclear
component.

The behavior  of the sources on  the plane formed by  the optical core
luminosity plotted vs.  the radio  core luminosity is related to their
optical  spectral classification.   Objects with  broad lines  show an
optical  excess  with  respect  to  the  correlation  found  for  FR~I
objects. This  excess is most  plausibly due to thermal  emission from
the accretion disk.

Only  one LEG  of  our  sample show  a  CCC, which  lies  on the  FR~I
correlation.   The  upper limits  for  the  other  two LEG  are  also
compatible with the correlation. This is essentially the same behavior
as observed in  low-z objects.  

The  two HEG  which have  a CCC,  as well  as the  five  optical upper
limits, are  consistent with the FR~I  correlation (within 2$\sigma$).
As  in the  case for  low-z  objects, the  nature of  their nuclei  is
classified  by  the  inclusion  of  a  further  parameter,  i.e.   the
luminosity of the [OIII] emission line.  The location of the nuclei in
the plane formed  by the equivalent width of  the [OIII] emission line
plotted vs.  the optical-to-radio core flux ratio indicates whether we
are directly observing the nuclear ionizing continuum or the source of
ionization is absorbed to our line-of-sight.

BLO  are  found  in  the  same region  as  their  low-z  counterparts,
therefore we interpret them as  unabsorbed nuclei.  The HEG lie in the
upper  part of  the plane,  indicating that  we are  observing  only a
fraction  of their ionizing  continuum source.   \object{3C~244.1} has
both  high EW([OIII]) and  optical excess,  therefore we  propose that
this source is a moderately absorbed BLO or, alternatively, a HEG with
a scattered  nucleus and with  a rather unusual  optical-to-radio core
flux  ratio.  Future deep  spectroscopic  observations  can test  this
hypothesis.

In  the FR~II--QSO  unification scenario  the number  ratio of  BLO to
narrow-lined  objects  has been  used  to  statistically estimate  the
opening angle  of the  obscuring torus.  In  the light of  our optical
studies  and  considering  the  most recent  spectral  information  we
confirm the  evidence that the opening angle  increases with redshift.
The most plausible explanation is provided by the receding torus model
(e.g. Lawrence~\cite{lawrence91}).

\begin{acknowledgements}

The authors thank William B. Sparks and Gabriele Giovannini for useful
discussions. This research has made use of the NASA/IPAC Extragalactic
Database  (NED) which is  operated by  the Jet  Propulsion Laboratory,
California Institute  of Technology, Under contract  with the National
Aeronautics  and Space  Administration. This  work has  been partially
supported by the  Summer Student Program (SSP) at  the Space Telescope
Science  Institute   and  by  contract   n.()  at  the   Institute  of
Radioastronomy.

\end{acknowledgements}

\end{document}